\documentclass{article}

\usepackage{PRIMEarxiv}

\usepackage[utf8]{inputenc} % allow utf-8 input
\usepackage[T1]{fontenc}    % use 8-bit T1 fonts
\usepackage{hyperref}       % hyperlinks
\usepackage{url}            % simple URL typesetting
\usepackage{booktabs}       % professional-quality tables
\usepackage{amsfonts}       % blackboard math symbols
\usepackage{nicefrac}       % compact symbols for 1/2, etc.
\usepackage{microtype}      % microtypography
\usepackage{lipsum}
\usepackage{fancyhdr}       % header
\usepackage{graphicx}       % graphics
\graphicspath{{media/}}     % organize your images and other figures under media/ folder
\usepackage{amsmath}

\title{Quasicrystalline Spin Foam with Matter: Definitions and Examples
}

\author{
  Marcelo Amaral, Richard Clawson and Klee Irwin\\
  Quantum Gravity Research \\
  Los Angeles, CA 90290, USA\\
  \texttt{Marcelo@QuantumGravityResearch.org}\\
  \texttt{Richard@QuantumGravityResearch.org}\\
  \texttt{Klee@QuantumGravityResearch.org}\\
}

\begin{document}
\maketitle

\begin{abstract}
In this work, we define quasicrystalline spin networks as a subspace within the standard Hilbert space of loop quantum gravity, effectively constraining the states to coherent states that align with quasicrystal geometry structures. We introduce quasicrystalline spin foam amplitudes, a variation of the EPRL spin foam model, in which the internal spin labels are constrained to correspond to the boundary data of quasicrystalline spin networks. Within this framework, the quasicrystalline spin foam amplitudes encode the dynamics of quantum geometries that exhibit aperiodic structures. Additionally, we investigate the coupling of fermions within the quasicrystalline spin foam amplitudes. We present calculations for three-dimensional examples and then explore the 600-cell construction, which is a fundamental component of the four-dimensional Elser-Sloane quasicrystal derived from the E8 root lattice.
\end{abstract}

% keywords can be removed

\tableofcontents

\section{Introduction}

The Feynman path integral for the gravitational field in the tetrad-connection formalism, which incorporates Dirac fermions and Yang-Mills gauge boson fields, is realized within the framework of loop quantum gravity. This is achieved through the use of spin foam models, which serve as a regularization of the formal path integral \cite{Rovelli-Vidotto-Book}. 

The EPRL spin foam model stands as the most extensively studied spin foam model to date, finding applications across a wide range of areas including cosmology and black hole physics. Notably, one of its significant achievements is the emergence of the Regge action in the semi-classical regime of large spins. Further details on this model can be found in \cite{Rovelli-Vidotto-Book} and the references provided therein.
The key idea underlying the construction of these spin foam models is to begin with general relativity as a topological theory with constraints. This theoretical framework enables the coupling of general relativity with fermions and bosons, facilitating the formulation of a formal path integral. By incorporating these different fields, spin foam models provide a comprehensive approach to studying the dynamics of gravity in a quantum framework. To regulate the path integral, a cellular decomposition dual to a triangulation of the original manifold is employed. This regularization procedure leads to the assignment of amplitudes to the cells of the decomposition. While much of the research in this area has focused on understanding the amplitude of specific cells, such as the vertex amplitude, there has been relatively less exploration into the behavior of the entire path integral as a whole.

To gain deeper insights into the spin foam model and explore aspects beyond the vertex amplitude, researchers have investigated simplified versions of the EPRL spin foam model. These simplified models allow for a better understanding of the theory's various building blocks.
For instance, one approach involves considering spin foam amplitudes with fixed boundary conditions, where the amplitudes sum over the bulk states. These bulk states can be constrained to match the boundary states, resulting in a simplified version known as the simplified EPRL model \cite{Speziale:2016axj}. It has been demonstrated that this simplified model successfully captures the key properties of the full EPRL model \cite{Dona:2019dkf}. Another noteworthy simplified model is the spin foam model with quantum cuboid intertwiners [\cite{Bahr:2015gxa}. In this model, the spin foam amplitudes are constrained to geometries compatible with coherent states selected on a cubic lattice. This constraint allows for a focused investigation into the interplay between the spin foam amplitudes and the specific geometric structure. These simplified models offer several advantages, such as improved tractability with numerical methods and direct consideration of states that are known to dominate the amplitude's behavior in the large spin limit \cite{Dona:2018nev,Gozzini:2021kbt}. This aligns with the claim that these simplified models capture the relevant features of the theory.

Motivated by the intriguing properties of quasicrystals \cite{BaakeGrimm,Moody2000mu,Senechal1995Book,Levine1986quasicrystals}, we propose a slightly more complex simplified model, while still maintaining tractability using similar techniques. Our approach begins by considering quasicrystalline spin networks as a novel subspace within the conventional Hilbert space of loop quantum gravity. This allows us to explore how these networks naturally impose constraints on the states of quantum geometry, selecting states based on the geometric properties exhibited by quasicrystals. Expanding upon this framework, we introduce quasicrystalline EPRL spin foam amplitudes, which are a modified version of the EPRL spin foam model. In this formulation, the internal spin labels are constrained to match the boundary data of quasicrystalline spin networks, which are represented by coherent states. This constraint ensures that the amplitudes capture the specific geometric properties associated with quasicrystals.

We begin by presenting the methodology employed to construct quasicrystalline spin networks and spin foam amplitudes. Subsequently, we conduct an analysis of specific three-dimensional examples to gain further insights. Furthermore, we extend our investigation to the more intricate 600-cell, which serves as a fundamental component of the four-dimensional Elser-Sloane quasicrystal derived from the E8 root lattice  \cite{ElserSloane1987,sadocmosseri1993}.
We found that in order to obtain the correct 3D quasicrystalline spin network boundary data for the amplitudes, when tetrahedra in the four-dimensional Elser-Sloane quasicrystal are rotated to align with the same three-dimensional space (hyperplane), a simple rotation exists from each tetrahedron's hyperplane to the target hyperplane. Notably, this rotation gives rise to a natural twist in the relationship between the tetrahedra in three dimensions \cite{FangRichard2018}. Importantly, this three-dimensional twist corresponds to the dual of the four-dimensional rotation between the hyperplanes. This intricate structural characteristic provides a unique and fertile platform for exploring the properties of quasicrystalline spin networks and their associated amplitudes in higher dimensions.

Furthermore, we explore the coupling of matter to the EPRL spin foam model, specifically focusing on the $SU(3)$ charge. This coupling can be interpreted within the context of the unifying gauge group E8 \cite{LisiLieE8,E8CarlosonTony}. By adopting this perspective, we gain valuable insights into the potential implications of quasicrystalline spin networks and spin foam amplitudes for understanding fundamental interactions. Moreover, it allows us to investigate the behavior of the amplitudes beyond a few cells, enabling us to calculate the amplitude for a complete quasicrystal tiling. All computations supporting the results presented in this paper are included in a companion Mathematica notebook, which is available on Wolfram's community platform \cite{qsnqsfnotebook2023}. These computations serve as a valuable resource for reproducing and verifying our findings.

The structure of this paper is as follows: In Section \ref{sec:definitions}, we provide the necessary definitions for quasicrystalline spin networks and quasicrystalline spin foams. In Section \ref{sec:computations},we delve into explicit examples of computations conducted in three dimensions, as well as the comprehensive model in four dimensions. This includes the construction of twisted boundary data and the incorporation of the fermionic sector. Finally, in Section \ref{sec:conclusion}, we present discussions and implications derived from our research.

\section{Quasicrystalline Spin Network and  Foam: Definitions}
\label{sec:definitions}

In the loop quantum gravity framework, the quantum states of gravity are represented by spin network states \cite{Rovelli-Vidotto-Book}. To understand how these states emerge, consider the following approach. Geometry is determined by the gravitational field, which is described by the tetrad field $e$ and a Lorentz connection $w$. This geometry can be triangulated with a desired level of precision. This process involves slicing the four-dimensional (4D) space into three-dimensional (3D) slices. In the boundary, the momentum conjugate to $w$ is an $sl(2,C)$-algebra-valued field $B$, where its electric and magnetic parts, $K$ and $L$, satisfy the so-called simplicity constraint $K=\gamma L$, where $\gamma$ is the Immirzi Parameter.

This approach motivates the use of a tetrahedron as a geometric building block, characterized by four normals that must satisfy the closure constraint. The quantization procedure suggests promoting geometric quantities, such as the normals, to operators acting within a Hilbert space. Consequently, the normals $\overrightarrow{n}_{a}$ should be described by quantum operators that account for gravity's quantum nature, subject to commutation relations. These relations can be postulated as:
\begin{equation}
[n_{a}^{i},n_{b}^{j}]=\beta\delta_{ab}\varepsilon_{k}^{ij}n_{a}^{k}\label{eq:normalsquantization}
\end{equation}
This proposition appears reasonable due to the invariance of geometric quantities under 3D rotations. This relation is essentially the $SU(2)$ one\footnote{One might anticipate $SL(2,C)$ symmetry, but the simplicity constraint allows for a mapping between $SL(2,C)$ and $SU(2)$. The quantum states of gravity are essentially $SU(2)$ spin networks. Representations of $SL(2,C)$ are labeled by a positive real number p and a non-negative half-integer k. However, due to the simplicity constraint, it is possible to show that in the large $j$ limit ($SU(2)$ quantum numbers), $p=\gamma k$, $k=j$, which selects $SU(2)$ subspaces.}, the double cover of $SO(3)$, assuring unitarity. Consider a triangle of the boundary tetrahedron and the tetrad field in the time-gauge where $e^{0}=dt$, $e^{i}=e_{a}^{i}dx^{a}$. Then a component of $\overrightarrow{n}_{a}$ can be written from the gravitational field
\begin{equation}
n_{a}^{i}=\frac{1}{2\gamma}\epsilon_{jk}^{i}\int_{t_{a}}e^{j}\wedge e^{k}.
\end{equation}
Thus, we observe that the postulated Eq. \ref{eq:normalsquantization} addresses the quantization of the gravitational field. The quantum gravity states are labeled by two quantum numbers, obtained from area and volume operators. The geometric area of a given triangle can be calculated as $A_{t_{a}}=\gamma|\overrightarrow{n}_{a}|$. And from the representation theory of $SU(2)$, one obtains Hilbert spaces $H_j$ and the spectrum $A{j}=\beta\sqrt{j(j+1)}$, where $\beta=8\pi\gamma\hbar G/c^{3}=\gamma A_{p}$, with $A_{p}$ being the Planck area.

Moreover, the volume operator can be expressed as
\begin{equation}
V_{n}=\frac{\sqrt{2}}{3}(\gamma A_{p})^{\frac{3}{2}}\sqrt{|\overrightarrow{n}_{1}.(\overrightarrow{n}_{2}\times\overrightarrow{n}_{3})|}
\end{equation}
This results in the Hilbert space being the tensor product of four $SU(2)$, $j$, representations meeting at the center of the tetrahedron., $\mathcal{H}_{j_{1}...j_{4}}=\textrm{Inv}_{SU(2)}(H_{1} \otimes ... \otimes H_{4})$, called the intertwiner space. 
The spectrum is obtained from 
\begin{equation}
V_{n}|\iota \rangle=\frac{\sqrt{2}}{3}(\gamma A_{p})^{\frac{3}{2}}\sqrt{|v|}|\iota \rangle
\end{equation}
where we examine the matrix elements of  $\langle \iota_{v}|\overrightarrow{n}_{1}.(\overrightarrow{n}_{2}\times\overrightarrow{n}_{3})|\iota_{v'} \rangle$, for which we have computed some examples in lower-dimensional spaces in \cite{qsnqsfnotebook2023}. The quantum gravity tetrahedron state can then be labeled by the two quantum numbers, $j$ and $v$. This extends over the entire 3D triangulated boundary as spin network states $|j_{l},v_{n} \rangle$ where $l$ represents the edges or links of the 2-complex dual to the triangulation and $n$ denotes the nodes. Observe the duality for the discretization: nodes $\leftrightarrow$ tetrahedra, edges $\leftrightarrow$ triangles.

Furthermore, we examine a coherent spin network as a spin network where the intertwiner states form coherent states, constructing tetrahedra coherent states. Initially, we consider $SU(2)$ coherent states
$|j,\overrightarrow{n} \rangle=D_{\overrightarrow{n}}(R)|j,j \rangle$, where $|j,j \rangle$
is the highest weight state of the $j$ representation and $D$ is the
Wigner matrix.
Consider an initial $|j,j \rangle$ state in the z direction, $z^{i}=(0,0,1)$, and define an $SO(3)$ rotation $R=e^{-i\phi n_{z}}e^{-i\theta n_{y}}$ by $R_{\overrightarrow{n}j}^{i}z^{i}=n^{i}$, where $\phi$ and $\theta$ are Euler angles labeling the rotations. The states $|j,\overrightarrow{n} \rangle$ form a family of states labeled by $\overrightarrow{n}$, which are coherent states. Now, consider states $|j_{1},\overrightarrow{n}_{1} \rangle\otimes|j_{2},\overrightarrow{n}_{2} \rangle \otimes|j_{3},\overrightarrow{n}_{3} \rangle \otimes|j_{4},\overrightarrow{n}_{4} \rangle$ projected to $\mathcal{H}_{j_{1}...j_{4}}$, $||j_{a},\overrightarrow{n}_{a} \rangle$. The coherent tetrahedra $||j{a},\overrightarrow{n}_{a} \rangle$ states are elements of the full spin network Hilbert space describing semi-classical tetrahedra.

Quasicrystals are structures similar to lattices but with tiles that repeat only aperiodically. Setting aside technical details for the moment, we can define \textbf{Quasicrystalline Spin Networks (QSN)} as the subset of spin network states found in the quasicrystal tiling geometry $\triangle$, $||j_{a},\overrightarrow{n}_{a} \rangle_{\triangle}$. So far, we have considered triangulations and their duals, so the quasicrystal structures need to be constructed in this way. Another option is to generalize the above construction to arbitrary discretization with nodes of arbitrary valence, which is already well understood \cite{Rovelli-Vidotto-Book}. For simplicity, we will stick with the simpler construction presented above.

Dynamics can be implemented through spin foam transition amplitudes, which assign individual amplitudes to vertices, edges, and faces of the 2-complex dual to the 4D triangulation. The dual map in 4D is as follows: : vertex $\leftrightarrow$ 4-simplex, edge $\leftrightarrow$ tetrahedron,
face $\leftrightarrow$ triangle. The amplitudes with fixed boundary states are given by
\begin{equation}
Z(j_{b},\overrightarrow{n}_{b})=\mathcal{N}\sum_{j_{f}/j_{b}}\prod_{f}A_{f}(j_{f})\prod_{v}A_{v}(j_{f},\overrightarrow{n}_{f})
\end{equation}
where $j_{f},\overrightarrow{n}_{f}$ 
represent the spin labels and normals associated with the faces of the 2-complex dual to the 4D triangulation. These quantities are allowed to vary within the bulk, where the dynamics of the spin foam model takes place. However, when a face is on the boundary, the spin labels and normals must be fixed to match the boundary data of the spin network states, denoted by $j_{b}$ and $\overrightarrow{n}_{b}$.
This constraint on the boundary ensures that the spin foam amplitudes are consistent with the coherent spin network boundary states, which represent the semi-classical geometry of the triangulated 3D boundary. By fixing the spin labels and normals on the boundary, the spin foam amplitudes capture the dynamics and evolution of the quantum geometry, while preserving the boundary conditions imposed by the spin network states. This approach is crucial for understanding the behavior of spin foam models and the interplay between the quantum geometry and the quasicrystalline structure.
Indeed, the specific form of the face and vertex amplitudes in spin foam models arises from discretizing the original gravitational action and employing the path integral formulation. In this discretization, the connection field associated with the gravitational field gives rise to group elements on the edges of the dual complex, while the tetrad field (which encodes the local geometry) contributes algebra elements on the faces. The choice of group and algebra elements depends on the dimensionality and the specific spin foam model being considered. For instance, in 3D the group elements are elements of the $SU(2)$ group. In the 4D case, the EPRL spin foam model employs $SL(2, C)$ elements, which is the complexification of the $SU(2)$ group and can be seen as a double cover of the Lorentz group $SO(3, 1)$. This choice of group elements accommodates both the gravitational field and its complex conjugate in the quantization procedure, while also being consistent with the simplicity constraint that relates $SL(2, C)$ and $SU(2)$ representations in the large spin limit. The specific form of the face and vertex amplitudes in the spin foam models encodes the dynamics of the quantum geometry and its evolution, providing a framework to study the interplay between quantum gravity and the underlying spacetime structure, such as the quasicrystalline geometries explored in this context.

A \textbf{Quasicrystalline Spin Foam (QSF)} transition amplitude is defined as a restriction to the spin foam amplitudes where the internal labels $j_f$ and $\overrightarrow{n}_f$ describe states that are associated with a quasicrystal geometry $\triangle$. This restriction is imposed such that the internal labels are consistent with the boundary quasicrystal geometry that has coherent states labeled by $j_b$ and $\overrightarrow{n}_b$. In this setting, the quasicrystalline spin foam amplitudes encode the dynamics of quantum geometries that exhibit aperiodic structures.

\subsection{Transitioning from Lattices to Quasicrystals: QSF Amplitudes}

In this section, we provide a brief overview of a potential implementation of quasicrystals, as found in the literature \cite{BaakeGrimm,Moody2000mu}.
The approach we examine is the cut-and-project scheme (CPS), which is represented by a 3-tuple $\mathcal{G}=\left(\mathbb{R}^{d},\mathbb{R}^{d'},\mathcal{L}\right)$. In this scheme, $\mathbb{R}^{d}$ denotes a real Euclidean space and $\mathcal{L}$ is a lattice within the space $\mathbb{R}^{d}\times \mathbb{R}^{d'}$. This lattice is often referred to as the mother lattice. 

The CPS involves two natural projections, $\pi$ and $\pi_{\bot}$, which map $\mathbb{R}^{d}\times \mathbb{R}^{d'}$ to $\mathbb{R}^{d}$ and $\mathbb{R}^{d'}$, respectively. 
These projections adhere to specific conditions, such as the injectivity of $\pi(\mathcal{L})$ and the density of $\pi_{\bot}(\mathcal{L})$ in $\mathbb{R}^{d'}$.
%These projections satisfy certain conditions, such as $\pi(\mathcal{L})$ being injective and $\pi_{\bot}(\mathcal{L})$ being dense in $\mathbb{R}^{d'}$.
The embedding space is represented as $\mathcal{E}=\mathbb{R}^{d}\times \mathbb{R}^{d'}$. In this representation, $\mathbb{R}^{d}$ is referred to as the parallel or physical space, while $\mathbb{R}^{d'}$ is considered the perpendicular or internal space. To proceed with this approach, a non-empty, relatively compact subset $K\subset \mathbb{R}^{d'}$ is required, which is known as the window. With $L=\pi(\mathcal{L})$ a well-defined map called the star map, $\star:L\rightarrow \mathbb{R}^{d'}$, is associated with a given CPS. This map is defined as $x\mapsto x^{\star}=\pi_{\bot}(\pi_{\mathcal{L}}^{-1}(x))$.

For a given CPS $\mathcal{G}$ and a window $K$, the quasicrystal point set can be generated by defining two additional parameters: a shift $\xi\in\mathbb{R}^{d}\times\mathbb{R}^{d'}/\mathcal{L}$, where $\xi{\bot}=\pi_{\bot}(\xi)$, and a scale parameter $\lambda\in\mathbb{R}$. The resulting projected set, denoted by $\triangle_{\xi}^{\lambda}(K)$, is referred to as a model set:
\begin{equation}
\triangle_{\xi}^{\lambda}(K)=\left\{ x\in L\mid x^{\star}\in\lambda K+\xi{\bot}\right\} =\left\{ \pi(y)\mid y\in\mathcal{L},\pi_{\bot}(y)\in\lambda K+\xi{\bot}\right\} \label{eq:modelset}
\end{equation}
Constructing a tiling on top of the quasicrystal point set is more intricate than with lattices. One possible approach to obtain a tiling in $\mathbb{R}^{d}$ is as follows: if two points in $L$ are connected in the mother lattice $\mathcal{L}$, they are also connected in $L$. Modifying $\xi$ or $\lambda$ can generate different tilings. A crucial property of quasicrystals relevant to our study is that for each model set $\triangle$ and associated tiling $\triangle_T$, with a total number of tiles $N$, there exists a finite number of tiles $t$, denoted as $N_t$, which repeat with an exact frequency, $f_t$, in the limit as $N\rightarrow\infty$.

We can express the general form of a QSF for a QSN boundary as follows:
\begin{equation}
Z(||j_{a},\overrightarrow{n}_{a}\rangle_{\triangle})=\mathcal{N}\sum_{j_{f}/j_{b}\subset\triangle_{\xi T}^{\lambda}}\prod_{f}A_{f}(j_{f})\prod_{v}A_{v}(j_{f},\overrightarrow{n}_{f})
\end{equation}
where $j_{f}/j_{b}\subset\triangle_{\xi}^{\lambda}{}_{T}$ denotes the constraint on the sum over states selected on the specific tiling configuration. In fact, another way to pose this is by noting that the method to alter tiling configurations is through changing $\xi$ and/or $\lambda$. As such, the sum goes over different tiling configurations $\triangle_{\xi T}^{\lambda}$, as discussed in the state sum presented in \cite{Amaral:2021}. We can isolate the amplitude for one configuration as follows:
\begin{equation}
Z(||j_{a},\overrightarrow{n}_{a}\rangle_{\triangle})=\mathcal{N}\sum_{j_{f}/j_{b}\subset\triangle_{\xi T}^{\lambda}}A_{\triangle_{\xi T}^{\lambda}}
\end{equation}
with
\begin{equation}
A_{\triangle_{\xi T}^{\lambda}}=\prod_{t_{f}}A_{f}(j_{f}){}^{Nf_{t_{f}}}\prod_{t_{v}}A_{v}(j_{f},\overrightarrow{n}_{f})^{Nf_{t_{v}}}\label{eq:onetilingprodAmp}
\end{equation}
where $t_{f}$ represents the different tile faces, which for quasicrystals have a finite number $N_{t_{f}}$, each with frequency $f_{t_{f}}$. Similarly, $t_{v}$ denotes the different tiles around a vertex, such as a tetrahedron or 4-simplex, which for quasicrystals have a finite number $N_{t_{v}}$, each with frequency $f_{t_{v}}$. Thus, computing the frequency of tiles in quasicrystals is one aspect of calculating QSF. In this paper, we will present some simple examples that avoid the detailed computation of frequency, which can be checked for specific quasicrystals in the literature \cite{BaakeGrimm}. In the following sections, we will present the results for computations of $A_{\triangle_{\xi T}^{\lambda}}$.

\section{Quasicrystalline Spin Network and  Foam: Examples}
\label{sec:computations}

In the following subsections, we present the results of addressing quasicrystalline spin foam amplitudes numerical computation for several examples in 3D and 4D. In 3D, we adapt the well-known Ponzano-Regge model for quasicrystals and implement QSF for the icosahedron and the octahedron, which are building blocks of known 3D quasicrystals. We show that in the large spin limit, it is in accordance with the expected power law behavior from analytical results, indicating that this kind of geometry dominates the amplitude.

In 4D, we study quasicrystalline EPRL spin foam amplitudes, particularly focusing on the 600-cell polytope, a building block of the more known 4D quasicrystal, the Elser-Sloane quasicrystal (ESQC) \cite{ElserSloane1987,sadocmosseri1993}. We demonstrate that the amplitude also exhibits the expected power law results for Euclidean boundary data. Subsequently, we couple fermionic cycles and compute the fermionic sector of the amplitude. We further investigate the boundary QSN data for the 600-cell, discussing the concept of parallel classes reduction \cite{FangRichard2018} within the states that dominate the amplitude in the large spin limit.

\subsection{3D Quasicrystalline Spin Foam}

For quantum gravity in three dimensions, we consider three-valent $SU(2)$ spin network states, and the 3D spin foam models reduce to the Ponzano-Regge model with $SU(2)$ symmetry \cite{Rovelli-Vidotto-Book}. The vertex amplitude is given simply by the {6j} symbol of $SU(2)$ representation theory. In the companion Mathematica notebook we reproduce the celebrated result that the vertex amplitude leads to Regge's action in the limit of large spin \cite{qsnqsfnotebook2023}. Constraining the model to a QSF, we obtain:
\begin{equation}
A_{\triangle_{\xi T}^{\lambda}}=\prod_{t_{f}}(-1){}^{Nf_{t_{f}}}(2j_{f}+1)^{Nf_{t_{f}}}\prod_{t_{v}}\{6j\}^{Nf_{t_{v}}}.
\end{equation}

Next, we present the computation for the icosahedron and octahedron.

\subsubsection{Amplitude for the Icosahedron}
\begin{figure}[!h]%{12cm}
	\centering{}
	\includegraphics[scale=0.5]{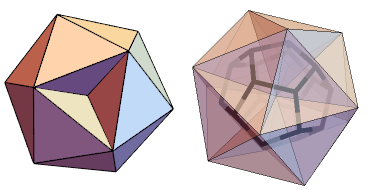}
 \caption{Left: Tetrahedra composing an icosahedron; Right: The dual dodecahedron.}
  \label{icosahimg}
\end{figure}

The boundary of the icosahedron consists of regular triangles. We insert a point in the center of the icosahedron to form tetrahedra, as depicted in Figure \ref{icosahimg}. 
Next, we insert a point inside each tetrahedron and connect them to obtain the dual 2-complex, which is a dodecahedron. As a result, there is one type of internal face, a pentagon, and one type of vertex tile, which is a tetrahedron where the boundary consists of a regular triangle and the non-regular internal ones are all the same. Thus, we have $Nf_{t_{v}}=20$ and $Nf_{t_{f}}=12$.

\begin{figure}[!h]%{12cm}
	\centering{}
	\includegraphics[scale=0.80]{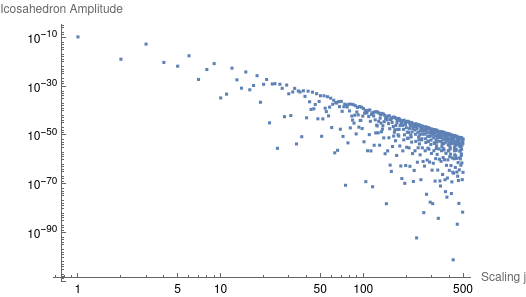}
 \caption{Asymptotic behavior of the icosahedron amplitude.}
  \label{icosahampasymfig}
\end{figure}

With two half-integer values, we cannot have the precise ratio between the boundary edge length and the bulk edge, so we approximate this ratio in the large j limit. The face of the dual complex is dual to an edge of the triangulation, so we can think of the j's being associated with the edge lengths. Essentially, we set one j to the other, scaled by the function Round of j. The {6j} symbol pattern follows the convention that the first row forms a triangle (in this case, the regular one), and the lower row has three edges coming from the equilateral triangle to the opposite vertex – in this case, the icosahedron center. The linear scaling in the log-log plot (indicating power law) is shown in Figure \ref{icosahampasymfig}.

\subsubsection{Amplitude for the Octahedron}
\begin{figure}[!h]%{12cm}
	\centering{}
	\includegraphics[scale=0.80]{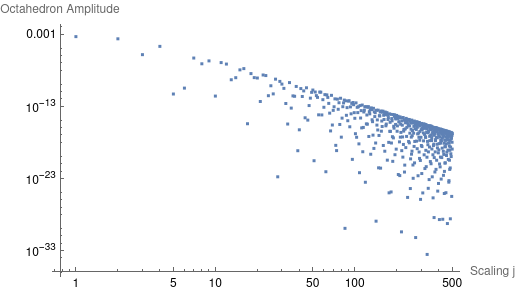}
 \caption{Asymptotic behavior of the octahedron amplitude.}
  \label{octahampasymfig}
\end{figure}

The amplitude computation for the octahedron proceeds in a similar manner as for the icosahedron. We insert a point in the center of the octahedron to triangulate it. The dual of this triangulation is a cube, which results in $Nf_{t_{v}}=8$ tetrahedra and $Nf_{t_{f}}=6$ square faces. The scaling of the amplitude is displayed in Figure \ref{octahampasymfig}.

\subsection{4D  Quasicrystalline Spin Foam}
\begin{figure}[!h]%{12cm}
	\centering{}
	\includegraphics[scale=0.80]{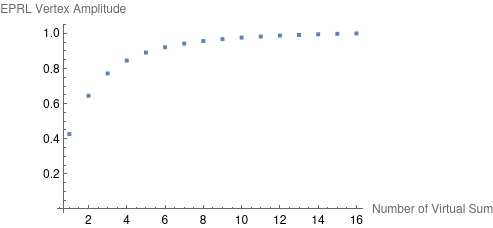}
  \caption{Convergence of the EPRL vertex amplitude with an increasing number of virtual sums.}
  \label{eprlampvirtualsumfig}
\end{figure}

For quantum gravity in four dimensions, we consider four-valent $SU(2)$ spin network states and the EPRL spin foam model \cite{Rovelli-Vidotto-Book}  with $SL(2,C)$ symmetry. In particular, we focus on the so-called simplified EPRL spin foam model, where the spins on the bulk virtual sums are fixed by the boundary spin network \cite{Speziale:2016axj,Dona:2019dkf,Bahr:2015gxa,Dona:2018nev,Gozzini:2021kbt}.
The face amplitude is given by the dimension of the $SU(2)$ j representation, $2 j_f + 1$.
The vertex amplitude is given by
\begin{equation}
A_{v}(j_{f},i_{e})=(\prod_{e=1...5}\sqrt{2i_{e}+1}\sum_{l_{f}k_{e}}(\prod_{e=2...5}(2k_{e}+1)B^{\gamma}(j_{f},i_{e};l_{f},k_{e}))\{15j\}(l_{f},k_{e}),\label{eq:EPRLveretxamplitude}
\end{equation}
where, in the simplified model, $l_{f}=j_{f}$, $k_{e}=i_{e}$, and there are no virtual sums. 
The \{15j\} symbol is the standard one of the first kind from $SU(2)$ representation theory. 
The so-called boost function $B^{\gamma}$ encodes the non-compact information of $SL(2,C)$.
The vertex amplitude can be computed efficiently using the spin foam code in C given in 
\cite{Gozzini:2021kbt}, which we used to compare with our implementation in Mathematica \cite{qsnqsfnotebook2023}, which works for our particular cases. For example, we reproduced the results of \cite[Table 3]{Speziale:2016axj} where for $\gamma=1.2$, $i_e = 0$ and equal spins $j=1,2,3$, $B^{\gamma}=$\{0.0236088, 0.00878174, 0.00485138\}. We also confirmed that the value of the amplitude converges quickly within the virtual sums, which supports the usefulness of the simplified model. See Figure  \ref{eprlampvirtualsumfig}.

\subsubsection{Amplitude for the 600-cell polytope}

In this study, we aim to elucidate the comprehensive structure of the ESQC amplitude by investigating its core component, the 600-cell. The 600-cell, a 4-dimensional hyper-regular polytope consisting of 600 tetrahedra with 20 meeting at each vertex, serves as the analogue of the icosahedron in 4D.
Recognized as a crucial component of the ESQC, the 600-cell has a significant influence on the shell structure, which stems from the quasicrystal's origin \cite{sadocmosseri1993}. Intriguingly, Sadoc and Mosseri's work presents an equation that suggests the number of points in each shell of the ESQC is divisible by 120 \cite{sadocmosseri1993}. However, they did not explicitly propose this divisibility as a conjecture in their paper and we found that their equation does not hold universally.
Despite this, based on the implications from their work and our own computations, we present a conjecture that has consistently held up: the number of points in each shell of the ESQC is divisible by 120. This pattern suggests the presence of at least one 600-cell within every shell. We have numerically corroborated this conjecture up to the 610th shell in our study.

To calculate the amplitude for the 600-cell, we adopt a similar methodology as employed for the 3D situation. 
 The EPRL QSF amplitude is given by
\begin{equation}
A_{\triangle_{\xi T}^{\lambda}}=\prod_{t_{f}}(2j_{f}+1)^{Nf_{t_{f}}}\prod_{t_{v}}A_{v}(j_{f},i_{e})^{Nf_{t_{v}}}
\end{equation}

The point set information for the 600-cell can indeed be obtained through the projection of the E8 root lattice using a cut-and-project scheme with $\mathcal{G}=\left(\mathbb{R}^{4},\mathbb{R}^{4'},E8\right)$. In this approach, only the first shell of points around the origin in E8, which corresponds to the 8-dimensional Gosset polytope, is considered.

Upon projecting the E8 root lattice, two 600-cells with different radii are produced. To select the point set for the larger 600-cell, a window can be derived that specifically targets the desired points. This technique, as described in \cite{ElserSloane1987,sadocmosseri1993} and \cite{qsnqsfnotebook2023}, successfully isolates the larger 600-cell's point set, providing the necessary information for further analysis of the 600-cell polytope and its implications in the context of quasicrystalline spin foam models.

The process for determining the amplitude is the same as the one used for the icosahedron. We begin by connecting the tetrahedral cells to the origin, creating 600 4-simplices. Next, we compute the dual 2-complex, which results in the 600-cell's dual polytope, the 120-cell. The 120-cell is a four-dimensional hyper-regular polytope consisting of 120 dodecahedra. In this case, there are $Nf_{t_{v}}=600$ vertices and $Nf_{t_{f}}=720$ pentagonal faces. When considering only the 20 tetrahedra (20G) surrounding a single vertex, we have $Nf_{t_{v}}=20$ vertices and $Nf_{t_{f}}=12$ faces.

We must then compute the vertex amplitude for the 600-cell 4-simplex. The spins are now mapped to the areas of triangles, which are dual to the faces. There are two types of triangles: a regular one on the boundary tetrahedra and another for the internal faces of the internal tetrahedra. A single 4-simplex consists of one boundary regular tetrahedron and four non-regular bulk tetrahedra. By fixing one spin and rounding the other while attempting to maintain the same ratio, we obtain the set of spins for the different areas: \{(1, 1), (1, 2), (2, 3), (2, 4), (3, 5), (3, 6), (4, 7), (5, 8), (5, 9), (6, 10)\}. 

It is important to note that the product in Eq. \ref{eq:EPRLveretxamplitude} covers only four tetrahedra, rather than all five in the 4-simplex. This technical detail is necessary to regularize the amplitude for $SL(2,C)$ \cite{Speziale:2016axj}. Coincidentally, the 4-simplex of the 600-cell contains only one regular tetrahedron, which is the boundary one. This is the ideal one to exclude from the integration, as it serves as the fixed boundary that determines the internal spins and intertwiner, while the sum is taken over the other four tetrahedra.

 With $\gamma = 0.153174$, the non-normalized amplitude for spins (1,2) is $-5.19684\times10^{-11}+7.23735\times10^{-25}i$. For spins (2,3) the amplitude is $1.48504\times10^{-12}-2.54231\times10^{-25}i$. For (2,4) the amplitude is $2.97604\times10^{-15}-5.84933\times10^{-29}i$. For
(3,5) the amplitude is $-5.81975\times10^{-16}+2.17437\times10^{-29}i$. And for (4,7) the amplitude is $2.89764\times10^{-19}+9.64954\times10^{-33}i$. For the 20G configuration with spins (1,2) the amplitude is $5.04001\times10^{-198}-1.40379\times10^{-210}i$.

\subsection{Twist: Boundary Coherent States from Dual Rotations}

Let us focus on one 4-simplex of the 600-cell. We fix this one to have the boundary tetrahedron normal in the time-gauge frame (-1,0,0,0). For each of the five 4-simplex tetrahedra, we couple a coherent tetrahedron state $||j_{a},\overrightarrow{n}_{a}\rangle$
\begin{align}
A_{v}(||j_{ab},\overrightarrow{n}_{ab}\rangle_{\triangle})= & (\prod_{a=1...5}\sqrt{2i_{a}+1}\sum_{l_{a}k_{a}}(\prod_{a=2...5}(2k_{a}+1)B^{\gamma}(j_{ab},i_{a};l_{a},k_{a}))\label{eq:EPRLveretxamplitudecoherent}\\
 & \{15j\}(l_{f},k_{e})\prod_{a=1...5}||j_{ab},\overrightarrow{n}_{ab}\rangle_{\triangle}.
\end{align}
Recall that a coherent tetrahedron is given explicitly by:
\begin{equation}
||j_{f},\overrightarrow{n}_{f}\rangle_{\triangle}=\sum_{m_{f}}\{4j\}\prod_{f=1}^{4}D_{m_{f},j_{f}}^{j_{f}}(\overrightarrow{n}_{f})
\end{equation}
with the Wigner matrix:
\begin{equation}
D^{j}(\overrightarrow{n})=e^{-i\phi J_{z}}e^{-i\theta J_{y}}e^{i\phi J_{z}}
\end{equation}
and the generalized Wigner \{4j\}  symbol, which is also used to compute the boost function, is given by:
\begin{equation}
\{4j\}(j_{f},m_{f},k)=\sum_{m}(-1)^{k-m}\times\{3j\}(j_{1},m_{1},j_{2},m_{2},k,m)\times\{3j\}(k,-m,j_{3},m_{3},j_{4},m_{4}).
\end{equation}
Here, the intertwiner number $k$ labels a valid representation on the tensor product space of the four face representations of the given tetrahedron. The challenge now is to find the precise set of Euler angles for our geometry of interest.

The boundaries of the QSF are networks of tetrahedra (QSN) described by the j's, which determine the areas of the triangles of each tetrahedron, as well as the normals of its triangles. More specifically, the amplitude's boundary data consists of areas and 3D normals in the time-gauge frame. Although a spin network could have arbitrary geometry in principle, the analysis of the asymptotic behavior of the amplitude in the literature indicates that Regge's geometry dominates the EPRL spin foam amplitude. To identify the $SO(4)$-invariant geometric area with the $SU(2)$ irreducible representations labeled by $j$, we must first find the transformation that brings all tetrahedra of our 4-simplex to the time-gauge and compute the resulting 3D normals for the faces. This leads to coherent states which have pairwise-opposite normals, implementing what we call the reduction of parallel classes:
\begin{equation}
\overrightarrow{n}_{ab}=-\overrightarrow{n}_{ba}\label{eq:normalsparallelclassesreduction}
\end{equation}
in addition to usual closure constraints:
\begin{equation}
\sum_{b\neq a}j_{ab}\overrightarrow{n}_{ab}=0.\label{eq:normalsclosure}
\end{equation}
We will observe that tetrahedra sharing faces in 4D will exhibit a twist in 3D. The transformations required to construct the 3D data from the 4D geometry involve rotations that bring all tetrahedra into the same 3D space, in addition to another rotation that implements Eq. \ref{eq:normalsparallelclassesreduction}. We have found that there are two dual rotations that satisfy the aforementioned conditions, and the angle involved is the 4D dihedral angle, which we will discuss further.

Let us examine how to obtain the 3D boundary data for one 4-simplex of the 600-cell. The five tetrahedra exist in different 3D spaces within 4D. First, we rotate them into the same 3D space as a reference tetrahedron in the time-gauge. To find the dual of a simple rotation in 4D, we identify the eigenplane of the rotation, take its dual plane, and construct a rotation with the same angle in that dual plane. The combination of the two rotations \cite{qsnqsfnotebook2023} generates the boundary network implementing Eq. \ref{eq:normalsparallelclassesreduction}.

Next, to compute the Euler angles, we interpret the normals as providing a coherent state that was initially in the z-axis \{0,0,1\} direction and then rotated to the general direction of the normal, allowing us to use the Mathematica function EulerAngles. The primary functions used to compute the coherent amplitude, as implemented in \cite{qsnqsfnotebook2023}, are rotMatrixToDual4D, CoherentTetrahedron, B4Simplified, and Wigner15jFk. We are able to compute the amplitude, yielding $7.40682\times10^{-17}-5.59056\times10^{-17}i$, which is at least 3 orders of magnitude larger than the amplitudes for random angle configurations, in agreement with the analytical result that this type of geometry is dominant. % The final data for the 4-simplex can be visualized in Figure \ref{4simplexto3DspinfoamTransFreedomFaceSharing1}.
% \begin{figure}[!h]%{12cm}
%	\centering{}
%	\includegraphics[scale=0.40]{data/E8toA4to4simplexto3DspinfoamTransFreedomFaceSharing1.png}
% \caption{600-cell 4-simplex  -- flattened and twisted by isoclinic rotations.}
% \label{4simplexto3DspinfoamTransFreedomFaceSharing1}
%\end{figure}

We can gain valuable insights into the general structure of the 600-cell amplitude by investigating its boundary QSN data. The boundary states involve only the regular tetrahedra. As we discussed earlier, it is possible to construct and visualize the 3D twisted tetrahedra for the boundary building block 20G around a single boundary vertex or starting from five tetrahedra sharing one edge. When tetrahedra in the 4D 600-cell are rotated to be in the same 3D space (hyperplane), a simple rotation exists from each tetrahedron's hyperplane to the target hyperplane. This rotation naturally induces a twisted relationship between the tetrahedra in 3D: the 3D twist is simply the dual to the 4D rotation between their hyperplanes. It's noteworthy that these rotations are isoclinic.
Moreover, the local structures discussed in this work can be comprehended within the context of a complete 3D quasicrystal, constructed from the ESQC using identical methods \cite{Fang2016xxx}.

\begin{figure}[!h]%{12cm}
	\centering{}
	\includegraphics[scale=0.40]{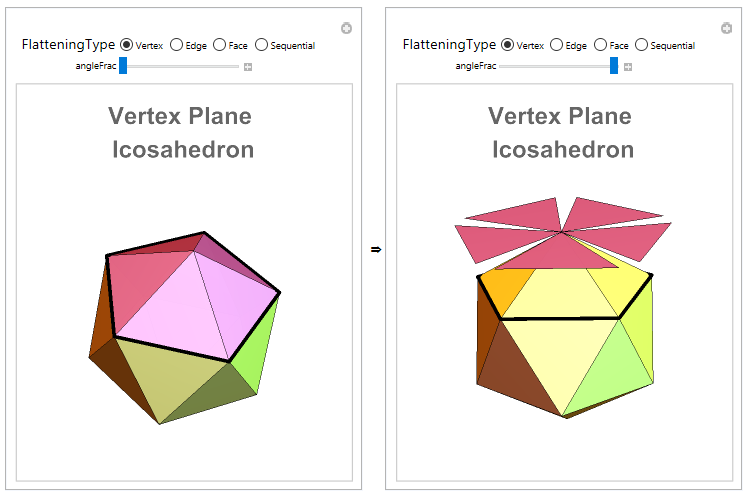}
  \caption{Icosahedron vertex cap centered on z-axis.}
  \label{icosahvertexcap}
\end{figure}

\begin{figure}[!h]%{12cm}
	\centering{}
	\includegraphics[scale=0.40]{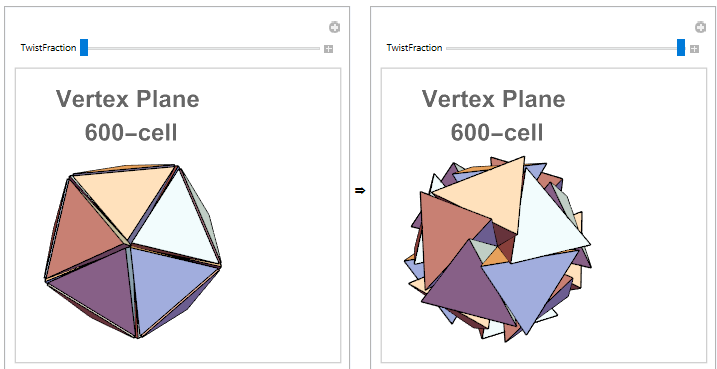}
  \caption{600-cell 20G center hyperplane 4Gs --flattened and twisted by isoclinic rotations.}
  \label{600cellvertexcaptwist}
\end{figure}

To draw a 3D analogy, let's consider one vertex on the icosahedron, which has 5 triangles around it, curved in different 2D spaces. In 3D, the boundary of these triangles forms a pentagon, which we call the vertex cap, as seen in Figure \ref{icosahvertexcap}.
Similarly, for the 20G in the 600-cell, the vertex cap is an icosahedron, and its dual dodecahedron is one dodecahedron of the dual 2-complex. We select one vertex in the time-gauge space to be the reference pole, then sort vertices by distance from the pole and determine the 3D space to be the target for the common rotation. Next, we construct rotation matrices for each cap tetrahedron, to rotate it directly to the vertex hyperplane defined as a simple rotation from the cell centroid to the pole (these are the normals of the cell's hyperplane and the vertex hyperplane).
The set of 20 rotated tetrahedra, all in the vertex hyperplane (but shifted, for plotting, down to the \{x, y, z\} hyperplane through the origin), are presented in Figure \ref{600cellvertexcaptwist} with the final twisted configuration.

\begin{figure}[!h]%{12cm}
	\centering{}
	\includegraphics[scale=0.40]{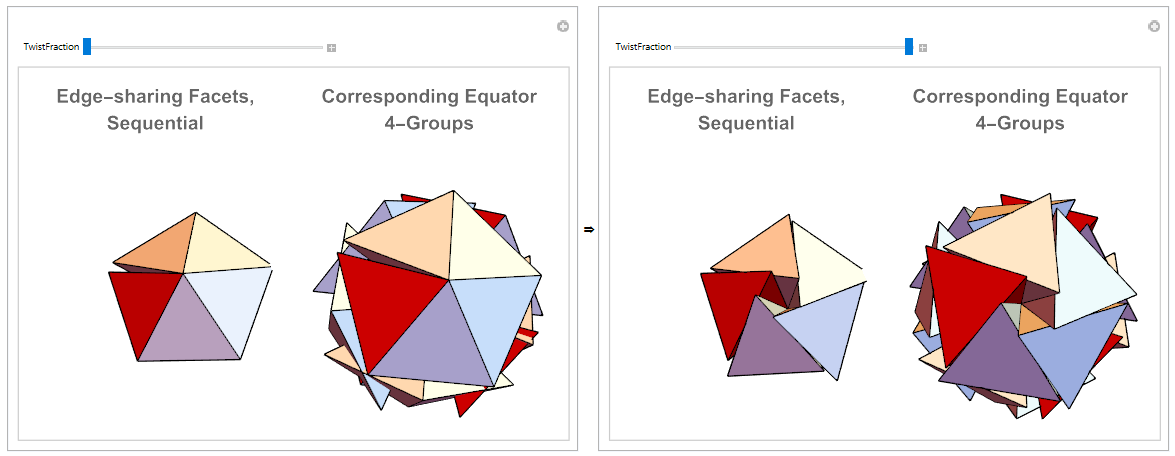}
 \caption{Flattening and twisting of 4Gs from 5 equatorial cuboctahedra of the 600-cell.}
  \label{twist20Gwithtetref}
\end{figure}

 It is also possible to perform the flattening and twisting for groups of 4 tetrahedra (4G) that share the same 3D space from 5 equatorial cuboctahedra of the 600-cell, with the 5 equators defined in various ways. Figure \ref{twist20Gwithtetref} illustrates the result where 4 red tetrahedra, which originally share the same 3D space, are kept fixed.

\subsection{Coupling fermions in the fundamental representation of SU(3)}

The process of discretizing fermions and Yang-Mills gauge bosons in triangulations \cite{Clemente:2021okd} or their dual \cite[Section 9]{Rovelli-Vidotto-Book} is similar to the standard lattice gauge theory procedures \cite{Rothe:1992nt}. In our case of interest, the goal is to attach group elements of the Lie group symmetry of the original action to the edges of the spin foam, thereby creating closed cycles around each face.
To obtain the full amplitude in the presence of matter, we need to couple fermionic and bosonic cycles with the previously considered amplitudes. The general form of this amplitude is given by
\begin{align}
Z(||j_{a},\overrightarrow{n}_{a},f_{a},b_{a}\rangle_{\triangle})= & \mathcal{N}\sum_{c/f_{a}b_{a}\subset\triangle_{\xi T}^{\lambda}}\sum_{j_{f}/j_{b}\subset\triangle_{\xi T}^{\lambda}}\prod_{f}A_{f}(j_{f})\prod_{v}A_{v}(||j_{ab},\overrightarrow{n}_{ab}\rangle_{\triangle})\nonumber \\
 & \int_{SU(3)}\prod_{c}A_{c}(j_{c},f_{c},b_{c}).\label{eq:amplitudewithsu3cycle}
\end{align}
The additional fermionic or bosonic Fock space information of the states is encoded by adding $f_{a}$ and $b_{a}$ to the boundary QSN coherent state.
The notation $c/f_{a}b_{a}\subset\triangle_{\xi T}^{\lambda}$ indicates that we need to sum over internal cycles $f_{c}$ and $b_{c}$ while keeping the boundary fermionic or bosonic states fixed (if there is a boundary edge).

It is natural to consider the minimal cycles over the edges of the dual 2-complex. In the case of the 600-cell, these are the pentagonal faces. For coupling the $SU(3)$ charge \cite{Amaral:2019rjb,AmaralChargeSpinNet2022}, we associate an $SL(2,C)$ and an $SU(3)$ group element, $U$ and $g$, respectively, to each half-edge. The value that enters $A_c$ is the trace in the fundamental representation for fermions ($\frac{1}{2}$ and (1,0)) and in the adjoint for bosons (1 and $(1,1)$).
For the EPRL spin foam model, let us consider a specific form of the fermionic cycles sector as given in \cite[Section 9]{Rovelli-Vidotto-Book}:
\begin{align}
\int_{SU(3)}dg_{ve}\prod_{c}Tr_{\frac{1}{2},(1,0)}\prod_{e\in c_{n}}(U_{ev}g_{ev}g_{ve}U_{ve})^{\varsigma^{ce}} & \times\\
\prod_{c}Tr_{1,(1,1)}\prod_{e\in c_{n}}(U_{ev}g_{ev}g_{ve}U_{ve})^{\varsigma^{ce}}.
\end{align}
For each regular tetrahedron of the QSN boundary state, there is an associated extra leg. Note that the dual of the edge is a tetrahedron in the 600-cell. Depending on the orientation of the cycle and the edge to match or not, $\varsigma^{ce}$ takes on a value of $\pm1$. In the large $j$, semi-classical limit, the amplitude is dominated by coherent states, such as the coherent tetrahedron discussed in the previous section, which is coupled to the matter group element $g$ here. However, we keep these states separated as the matter elements are taken in a precise lower representation, while the gravitational part fluctuates over all representations.

There are integrations over the group elements for each edge, which can have more than one cycle going over it. For instance, the 120-cell has 3 dodecahedra around one edge, and thus 3 pentagonal faces sharing one edge. Fermionic properties restrict us to consider only 2 cycles per pentagon (the 2 different orientations). We can have integrations with mostly 6 group elements (the trace in the fundamental representation, which are the characters $\chi$). Let us consider only one orientation and the $SU(3)$ integrations. We further simplify by considering just one element per edge (the product of the elements on the half edges) and have the integration for the fermionic sector
\begin{equation}
I_{1}=\int_{SU(3)}dg_{1}\chi(g_{1}G_{1})\chi(g_{1}G_{2})\chi(g_{1}G_{3})
\end{equation}
where the measure $dg_{1}$ is the $SU(3)$ Haar measure and $G_{i}$
encodes the product of the other edges' group elements in the cycles. The characters $\chi(g)$ are given by the trace in the fundamental representation, which is a number that depends on the number of cycles going over one edge.

\begin{figure}[!h]%{12cm}
	\centering{}
	\includegraphics[scale=0.80]{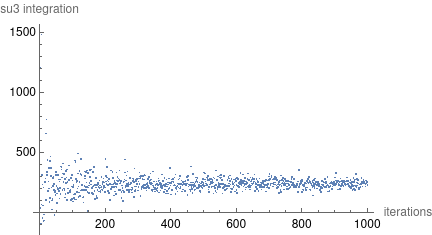}
  \caption{Approximating SU(3) integration of a fermionic cycle amplitude.}
  \label{su3integration}
\end{figure}

The integration can be approximated by Monte Carlo methods, which involve generating a large set of random $SU(3)$ matrices in the fundamental representation and computing the trace, as shown in Figure \ref{su3integration}.

The situation is similar for the $SL(2,C)$ group elements $U$ in the fundamental representation. It is interesting to note that the $j$ values are associated with the triangle areas of the tetrahedron dual of the edges. While the $SL(2,C)$ group elements of gravity can be in any representation, the fermion is in the 1/2 representation. This indicates that adding one fermionic family of cycles adds an additional layer with small areas compared with the large $j$ limit of the gravitational sector.

Usually, one would consider that the $SL(2,C)$ $U$'s in the fundamental representation and general $SL(2,C)$ gravitational elements on the gravitational sector would be the same to couple fermions with gravity and integrate over everything. However, we keep the two sectors' integrations separated. The gravitational amplitude was computed in the previous section. We note that at this constrained, simplified model level, bosonic and fermionic cycles are of the same nature; only the specified value of the integral changes. Thus, what matters are the overlaps between different cycles, which would increase the value of the amplitude going over one edge of the dual 2-complex.

The state at each tetrahedron is labeled by the coherent state spin network with the spins, intertwiners, and normals. We now add the fermionic Fock space labeled by $f_b$, which can be 0 for the vacuum. If there is a matrix $U$ in the fundamental representation, it means that there are two states labeled by $\pm 1/2$ spins. There can also be two cycles or two matrices, in which case the tensor product space has a new quantum number.

Dimensional analysis indicates that there is a hidden volume of the tetrahedron in the fermionic amplitude \cite[Section 9]{Rovelli-Vidotto-Book}, suggesting that charge quantum numbers may be derived from the geometry, much like with the spin quantum numbers. For the sake of this discussion, let us assume that these QSF amplitudes are the deepest reality possible at the Planck scale. Then we can assume that any value of any field must be explained by this spin foam level or some emergent level on top of it. If a field has some value at some spacetime point, this value is part of reality and must be given from the reality structure, which, upon our assumptions, is this network.

There are two ways to assign a value to a fundamental field from the network structure. The first is randomness, which is similar to the Monte Carlo method used, where we integrate over all possible values. Alternatively, we can get the values from the geometric structure implied by the given network, such as connecting the volume with the charge. This approach would provide a precise value, and the integration would not be necessary.

It is important to note that as the gravitational sector of the amplitude drops exponentially with the number of tiles, a local fermionic cycle dominates. The minimum fermionic cycle consists of one pentagon going over 5 4-simplices. The boundary state is given by 5 boundary regular tetrahedra, which we can think of as representing the transition between the 5 coherent tetrahedra states.

\subsection{Note on E8 Unification}

The ESQC emerges as a projection from the E8 root lattice, which is self-dual and associated with the E8 Lie algebra and Lie group. In the representation theory of Lie algebras, possessing both the root and dual weight lattices enables the recovery of the primary structure of the algebra and the group manifold. It is noteworthy to observe how the information on the quasicrystalline spin network considered here can be mapped to the E8 lattice.

It is essential to highlight that the matter cycles contributing to the general amplitude \ref{eq:amplitudewithsu3cycle} are derived by starting with an action for Einstein-Weyl-Yang-Mills-Dirac theory as discussed in \cite{Rovelli-Vidotto-Book}. The concept of grand unification theories (GUT) involves constructing the action from a larger unification field that encompasses the standard model fields and gravity, leading to the inclusion of more Yang-Mills and Dirac terms in the original action. One approach considers a single high-dimensional group of symmetry with all fields stemming from the E8 Lie group \cite{LisiLieE8, Chester:2020nim}. Subsequently, it is customary to divide it to separate the gravitational symmetry sector from the matter sector, with the matter represented by higher-dimensional Yang-Mills and Dirac terms. Typical unification groups for the standard model are $SU(5)$ and $SO(10)$ \cite{gilmore}. E8 can be broken down into an $SO(3,1)$ Lorentz group plus a set of matter unification groups, potentially accounting for the known three fermionic families. For instance, one can obtain the Lie algebras associated with $SO(10)$ or $SU(5)$ in both possible signatures of $\mathfrak{so}_{12,4} \subset \mathfrak{e}_{8(-24)}$, where $\mathfrak{e}_{8(-24)}$ is the noncompact (quaternionic) real form of the Lie algebra $\mathfrak{e}_8$ \cite{Chester:2020nim}. For example: $\mathfrak{e}_{8(-24)}\rightarrow\mathfrak{so}_{4,12}\rightarrow\mathfrak{so}_{10}\oplus\mathfrak{so}_{4,2}\rightarrow\mathfrak{su}_{5}\oplus\mathfrak{so}_{4,2}\oplus\mathfrak{u}_{1}$. By decomposing the root polytope into sub-root-polytopes of sub-algebras contained within it, the root lattice can be employed to obtain the sub-algebras. As a lower-dimensional example, consider the $SU(3)$ group and associated Lie algebra, which possess root and weight lattices that are dual A2 lattices. The root polytope, which provides the adjoint representation, is a hexagon, and the two fundamental representations are represented by triangles. The unification group $SU(5)$ has an associated Lie algebra with the root polytope being the rectified 5-cell, composed of tetrahedra and octahedra whose faces are all equilateral triangles. This allows for the recovery of the $SU(3)$ sub-algebra and groups from $SU(5)$ in a geometric manner within the root system \cite{Koca4}. For E8, the root polytope is the well-known Gosset polytope. E8 has 248 generators, of which 8 are the standard Cartans defining the 8D Cartan sub-algebra, and 240 are the root vectors. The Gosset root polytope is formed by the 240 root vectors. The ESQC is a cut-and-project scheme for the E8 lattice, and the 600-cell building block emerges as a projection of the Gosset polytope. In this context, it is possible, in principle, to isolate the lower-dimensional root polytopes in the Gosset and track the information to the 600-cell to map the group elements for $SL(2,C)$ and $SU(3)$ that we have considered.

In this setting, the ESQC serves as a bridge between the high-dimensional E8 lattice and the lower-dimensional structures we are interested in. By analyzing the geometric relationships between the root polytopes and their projections, we can gain insights into how the various group elements and algebraic structures emerge in the quasicrystalline spin network. This approach allows us to establish connections between the high-dimensional symmetries of the E8 Lie group and the lower-dimensional $SL(2,C)$ and $SU(3)$ group elements that we encounter in the spin foam amplitudes.

To illustrate this, let us consider the simplest sub-Lie-algebra and group $SU(2)$. For each non-zero pair of root vectors, $\pm \alpha$, there is an $SU(2)$ subalgebra with generators:
\begin{align}
E^{\pm}= & |\alpha|^{-1}E_{\pm\alpha}\nonumber \\
E_{3}= & |\alpha|^{-2}\alpha.H
\end{align}
with $[E_{\alpha},E_{-\alpha}]=\alpha.H$. %( $[E_{3},E^{\pm}]=\pm E^{\pm}$, $[E^{+},E^{-}]=E_{3}$)

The highest weight state $|j,j\rangle$ of representation $j$, which can give the $j$ label for triangle areas or describe a coherent state on the quasicrystal geometry, is linked to $\alpha$ by the equation:

\begin{equation}
\alpha.H|j,j\rangle=|\alpha|^2j|j,j\rangle.
\end{equation}

Thus, we can consider that the Hilbert spaces labeled by $j$ in the spin network are coming from E8 representation spaces. The one-dimensional lattice given by $\alpha$ is the A1 root lattice embedded in higher dimensions, and the root lattice provides the spin quantum numbers from the root vectors. Further analyses of the 600-amplitude as encoding GUT information have yet to be conducted, but this framework offers a foundation for understanding how these high-dimensional algebraic structures manifest in the quasicrystalline spin network and impact the calculation of amplitudes.

\section{Conclusion}
\label{sec:conclusion}
In conclusion, our study has provided valuable insights into the relationship between quasicrystals, spin networks, and the EPRL spin foam model. This work opens up avenues for a deeper understanding of quantum geometry and the unification of fundamental forces.

In this paper, we have presented quantum gravity amplitudes within a spin foam model that is constrained to quasicrystal discrete geometry. By fixing the boundary coherent states to be picked in the quasicrystal geometry, the computed amplitudes can be interpreted as potential observables or transitions for quantum states that exhibit aperiodic structures. These observables possess both matter and geometric quantum numbers, making them significant for future experimental testing and further exploration of the fundamental nature of the universe.

\begin{figure}[!h]%{12cm}
	\centering{}
	\includegraphics[scale=0.50]{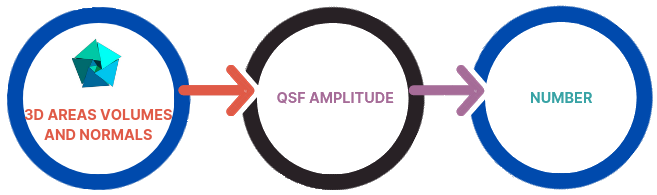}
 \caption{Flow chart for the amplitude. The full flow process don't address translations with regard the 3D data.}
  \label{flowchartamplitude1}
\end{figure}

The quasicrystalline spin foam amplitude can be viewed as a mathematical function that takes specific input data and generates a complex number, as illustrated in Figure \ref{flowchartamplitude1}. This function encapsulates the dynamics of the system and provides a mathematical representation of the probability amplitude associated with a given spin foam configuration.
Understanding the behavior of this function is crucial for developing a deeper comprehension of quantum gravity and its relationship with other fundamental forces. Our study has shed light on the mathematical structure of this function, particularly in the context of quasicrystal discrete geometry and its interplay with spin networks and the EPRL spin foam model. Notably, we have explored the possibility of computing the amplitude for a tiling configuration, as expressed in Eq. \ref{eq:onetilingprodAmp}. However, a complete understanding of quasicrystalline spin network evolution would require consideration of different tiling configurations within the full sum.

Various analyses conducted in the literature have provided insights into the constraints on the geometric interpretation of the boundary data. However, it is important to note that the interpretation of these constraints remains a topic of ongoing debate. One of the key constraints that has been identified in the literature is represented by Eqs. \ref{eq:normalsparallelclassesreduction} and \ref{eq:normalsclosure}. In the conventional approach, one typically begins with arbitrary 3D data and proceeds to reconstruct the implied 4D geometry. However, in our study, we adopted a different approach by starting with a regular 4D geometry, such as the 600-cell. From this initial 4D geometry, we constructed the 3D boundary data by employing a specific set of dual 4D rotations. This process ultimately led to the formation of twisted set of tetrahedra, which are grouped together based on shared faces.
These findings highlight the intricate relationship between the 4D geometry and the resulting 3D boundary data, shedding light on the role of dual 4D rotations in shaping the geometric properties of the system. As further research is conducted, it will be important to continue exploring and refining our understanding of these constraints, contributing to the ongoing discourse surrounding the interpretation of the boundary spin network data.

These structures are understood to exist in the time-gauge frame of reference, which intriguingly aligns with the frame typically employed to define fermions. However, we have not encountered any constraints on translations within this frame of reference, indicating the need for further investigation regarding Poincaré invariance and related considerations. As we continue to explore the interplay between quasicrystals, spin networks, and the EPRL spin foam model, we anticipate that new insights will emerge, providing deeper understanding of the fundamental nature of the universe.

\section*{Acknowledgments}
We are sincerely grateful to Raymond Aschheim, David Chester, and Fang Fang for their invaluable discussions on the subject of quasicrystalline spin networks and spin foams.

%Bibliography
%\bibliographystyle{unsrt}  
%\bibliography{references}  

\begin{thebibliography}{999}
%

\bibitem{Rovelli-Vidotto-Book} 
	Rovelli, C., Vidotto, F.:
	Covariant Loop Quantum Gravity.
	  Cambridge University Press  1 edition, (2014).

\bibitem{Speziale:2016axj}
    Speziale, S. Boosting Wigner\textquoteright{}s nj-symbols,
    J. Math. Phys. \textbf{58} (2017) no.3, 032501.
    [arXiv:1609.01632 [gr-qc]].

\bibitem{Dona:2019dkf}
    Don\`a, P., Fanizza, M., Sarno, G. and Speziale, S.
    Numerical study of the Lorentzian Engle-Pereira-Rovelli-Livine spin foam amplitude,
    Phys. Rev. D \textbf{100} (2019) no.10, 106003
    [arXiv:1903.12624 [gr-qc]].

\bibitem{Bahr:2015gxa}
    Bahr, B. and Steinhaus, S.
    Investigation of the Spinfoam Path integral with Quantum Cuboid Intertwiners,
    Phys. Rev. D \textbf{93} (2016) no.10, 104029
    [arXiv:1508.07961 [gr-qc]].

\bibitem{Dona:2018nev}
    Dona, P. and Sarno, G.
    Numerical methods for EPRL spin foam transition amplitudes and Lorentzian recoupling theory,
    Gen. Rel. Grav. \textbf{50} (2018), 127.
    [arXiv:1807.03066 [gr-qc]].

\bibitem{Gozzini:2021kbt}
    Gozzini, F.
    A high-performance code for EPRL spin foam amplitudes,
    Class. Quant. Grav. \textbf{38} (2021) no.22, 225010
    [arXiv:2107.13952 [gr-qc]].

\bibitem{BaakeGrimm} 
 	Baake, M. and Grimm, U. Aperiodic Order, Cambridge University Press, (2013).

\bibitem{Moody2000mu}
    Moody, R. V.
    Model sets: A Survey, (2000).
    [arXiv:math/0002020 [math.MG]].

\bibitem{Senechal1995Book}
    Senechal, M.~J., {\em Quasicrystals and Geometry}, Cambridge University Press, (1995).

\bibitem{Levine1986quasicrystals}
    Levine, D. and  Steinhardt, P.~J., Quasicrystals. I. Definition and structure. {\em Phys Review B}, {\bf 1986}, {\em 34}, 596, (1986). 

\bibitem{ElserSloane1987}
    Elser, V. and Sloane, N. J. A., A highly symmetric four-dimensional quasicrystal, J. Phys. A 20, 6161, (1987).

\bibitem{sadocmosseri1993}
    Sadoc, J. F. and Mosseri, R. The E8 lattice and quasicrystals: geometry, number theory and quasicrystals. J. Phys. A: Math. Gen. 26 1789, (1993).
    
 \bibitem{FangRichard2018}
    Fang, F., Clawson, R. and Irwin, K. Closing Gaps in Geometrically Frustrated Symmetric Clusters: Local Equivalence between Discrete Curvature and Twist Transformations. Mathematics 2018, 6, 89.
    
\bibitem{LisiLieE8} 
    Lisi, A. G. An Exceptionally Simple Theory of Everything.  arXiv:0711.0770 [hep-th]. (2007).

 \bibitem{E8CarlosonTony} 
   Castro. C. A Clifford algebra-based grand unification program of gravity and the Standard Model: a review study.
   Canadian Journal of Physics, 92(12): 1501-1527. (2014).
   
\bibitem{qsnqsfnotebook2023} 
    Amaral, M., Clawson, R. Quasicrystalline spin foam with matter: computations.
    Wolfram Community, STAFF PICKS, June 1, (2023).
    \url{https://community.wolfram.com/groups/-/m/t/2929437}.

\bibitem{Amaral:2021}
    Amaral, M., Fang, F., Aschheim, R. and Irwin, K. On the Emergence of Spacetime and Matter from Model Sets. Preprints.org 2021, 2021110359. \url{https://doi.org/10.20944/preprints202111.0359.v2}.

\bibitem{Fang2016xxx} 
	Fang, F. and Irwin, K. An Icosahedral Quasicrystal and E8 derived quasicrystals. 
	arXiv:1511.07786v2  [math.MG]. (2016).
 
\bibitem{Clemente:2021okd}
    Clemente, G., Candido, A., D'Elia, M. and Rottoli, F.
    Coupling Yang--Mills with Causal Dynamical Triangulations,
    PoS \textbf{LATTICE2021}, 254 (2022)
    [arXiv:2112.03157 [hep-lat]].

\bibitem{Rothe:1992nt}
    Rothe, H.~J.
    Lattice Gauge Theories : An Introduction (Fourth Edition),
    World Sci. Lect. Notes Phys. \textbf{43}, 1-381 (1992)
    World Scientific Publishing Company, 2012.

\bibitem{Amaral:2019rjb}
    Amaral, M., Aschheim, R., and Irwin, K.
    Quantum gravity at the fifth root of unity,
    Phys. Open \textbf{10}, 100098 (2022).
    [arXiv:1903.10851 [hep-th]].

\bibitem{AmaralChargeSpinNet2022} 
    Amaral, M. On charge-spin networks from multiway systems branchial graphs. Wolfram Community, STAFF PICKS, (2022).
    \url{https://community.wolfram.com/groups/-/m/t/2311181}.

\bibitem{Chester:2020nim}
    Chester, D. Marrani, A. and Rios, M.
    Beyond the Standard Model with Six-Dimensional Spinors,
    Particles \textbf{6}, no.1, 144-172 (2023)
    [arXiv:2002.02391 [physics.gen-ph]].

\bibitem{gilmore} 	
	Gilmore, R. Lie groups, Lie algebras, and some of their applications.
	Dover publications, INC. (2005).	
 
\bibitem{Koca4} 
	Koca M., Koca O. N. and Al-Siyabi, A.,
	SU(5) Grand Unified Theory, its Polytopes and 5-fold Symmetric Aperiodic Tiling.
 	International Journal of Geometric Methods in Modern Physics (IJGMMP), Volume.15 (4) (2018).
 	

\end{thebibliography}

\end{document}